\documentclass[pra,twocolumn,showpacs,superscriptaddress,floatfix]{revtex4-2}
\usepackage{graphicx}
\usepackage{epstopdf} 
\usepackage{amsmath}
\usepackage{epsfig}
\usepackage{latexsym}
\usepackage{amsfonts}
\usepackage{graphics}
\usepackage{bm}
\usepackage{braket}
\usepackage{dsfont}
\usepackage{soul}
\usepackage{ulem}
\usepackage{bbm}
\usepackage{hyperref}

\usepackage{mathtools}

\usepackage{helvet}

\graphicspath{ {./Graphics/} }
\begin{document}

\date{\today}
\author{J. Mumford}
\affiliation{Homer L. Dodge Department of Physics and Astronomy, The University of Oklahoma, Norman, Oklahoma 73019, USA}
\affiliation{Center for Quantum Research and Technology, The University of Oklahoma, Norman, Oklahoma 73019, USA}
\affiliation{Department of Physics and Astronomy, University of Victoria, Victoria, British Columbia V8P 5C2, Canada}
\author{R. J. Lewis-Swan}
\affiliation{Homer L. Dodge Department of Physics and Astronomy, The University of Oklahoma, Norman, Oklahoma 73019, USA}
\affiliation{Center for Quantum Research and Technology, The University of Oklahoma, Norman, Oklahoma 73019, USA}

\title{Universal Long-Time Behavior of the Quantum Fisher Information in Dynamical Quantum Phase Transitions}

\begin{abstract}
We investigate dynamical quantum phase transitions (DQPTs) in quantum systems that possess well-defined classical limits, focusing on the spinor Bose-Einstein condensate and the Lipkin-Meshkov-Glick model. We diagnose the DQPTs with the long-time average of the quantum Fisher information (QFI) showing that it abruptly changes at the transition point.  Using mean-field and semiclassical approximations, we demonstrate that the long-time average of the QFI reveals universal behavior that persists across different systems.
\end{abstract}

\maketitle

\section{Introduction}

Recent advancements in experimental techniques have enabled the isolation and control of quantum systems at the level of individual particles, opening the door to detailed investigations of quantum many-body dynamics far from equilibrium. In this context, dynamical quantum phase transitions (DQPTs) \cite{schiro10, diehl10,schiro11,sciolla11,sciolla13,vosk14,halimeh17,zunkovic18,marino22} have emerged as a topic of significant interest, as they offer insight into how the long-time behavior of quantum systems undergo sudden changes as external parameters are changed. Typically, DQPTs are probed by performing a quench of the system, after which the transition manifests as a nonanalytic change in the long-time average (LTA) of an observable which acts as an order parameter.  This approach mirrors equilibrium quantum phase transitions which are characterized by discontinuities in the expectation value of an observable at equilibrium.  It is important to distinguish this type of DQPT from another where the transition is instead characterized by nonanalytic behavior in the dynamics itself \cite{zvyagin16,heyl18}.  
Advances in the ability to sustain coherent dynamics over long timescales and in state preparation has led to DQPTs being detected in a wide range of experimental platforms, including ultracold atomic gases \cite{schmaljohann04,yang19,muniz20,tian20}, trapped ions \cite{zhang17,smale19}, superconducting qubits \cite{xu20}, and nuclear spin ensembles \cite{jun17,nie20}. 


%

DQPTs provide a powerful lens for exploring connections between different aspects of quantum dynamics.  A common approach is to study systems with well-defined classical limits, in which case the dynamics can be described in terms of a fictitious classical particle moving in an effective one-dimensional potential \cite{swan21,guan25}.  In this picture, a DQPT often corresponds to the particle gaining enough energy to overcome some barrier, so it can explore more of the potential.  The top of the barrier, often referred to as an unstable fixed point or separatrix, marks the boundary between distinct dynamical behavior.  It is this point as well as the region around it which links many of the seemingly disparate topics in quantum dynamics. For instance, dynamics near the barrier can lead to rapid information scrambling, which has been shown to enhance quantum sensing \cite{guan21,munoz23,zhou23,zhang24}. Moreover, the dynamics in this region often resemble the exponential growth characteristic of classical chaos, motivating ongoing efforts to distinguish the effects of chaotic behavior from those arising purely from unstable fixed points \cite{pilatowsky20,xu20b,hashimoto20,bhattacharjee22}.

A versatile tool for probing nonequilibrium quantum dynamics is the quantum Fisher information (QFI) \cite{helstrom69,toth14,pezze18} which is a fundamental concept in quantum information theory \cite{braunstein94,petz11,hyllus12,toth12,yadin16}.  It quantifies how much information about a classical parameter $\phi$ is encoded in a quantum state $\hat{\rho}(\phi)$, and it sets a limit on how precisely that parameter can be estimated from measurements \cite{braunstein94}.  In quantum metrology, the QFI is commonly used as a probe of multipartite entanglement \cite{hyllus12,toth12}, however, recently it was shown that the QFI is directly related to out-of-time-ordered correlators \cite{garttner17,garttner18} which are used to explore the relations between chaos, thermalization, criticality and information scrambling \cite{swan19,kidd20,swan20,kirkova22}. In parallel, the QFI has also emerged as a diagnostic of DQPTs \cite{lerose20,guan21,munoz23}, further broadening its scope and applicability.

In this paper, we investigate DQPTs in quantum systems that possess well-defined classical limits, specifically focusing on the spinor Bose-Einstein condensate (BEC) \cite{kawaguchi12,ueda17} and the Lipkin-Meshkov-Glick (LMG) \cite{lipkin65,meshkov65,glick65} model. To characterize DQPTs in these systems, we compute the LTA of the QFI.  The observable that encodes the information of $\phi$ in the state is chosen so that its classical counterpart corresponds to the position variable of a classical particle in an effective potential. This connection enables us to interpret the DQPTs as arising from sudden changes in the underlying potential. We analyze two distinct types of DQPTs: one involving a sudden change in the shape of a single-well, and another in which the particle transitions from being trapped to untrapped in a symmetric double-well potential. The classical interpretation provides a basis for deriving general analytic expressions for the LTA of the QFI, which we use to quantitatively diagnose the DQPT separating the two dynamical phases. The methods we use in this work are broadly applicable and can be extended to other types of effective potentials.  

\section{Preliminary Approximations\label{sec:Sapprox}}

\subsection{Long-Time Average of the Quantum Fisher Information}

In interferometric measurements, it is common to encode a parameter $\phi$ into a state via a unitary operator $\hat{\rho}(\phi,t)=e^{i \phi \hat{X}} \hat{\rho}(t)e^{-i \phi \hat{X}}$ \cite{cronin09} where $\hat{X}$ is an observable.  When making measurements on a pure state, $\hat{\rho}(t) = \vert \Psi(t)\rangle\langle \Psi(t)\vert$, the QFI is expressed in terms of the variance of $\hat{X}$,

\begin{equation}
F_Q(t) = 4 \left [\langle \Psi(t) \vert \hat{X}^2 \vert \Psi(t) \rangle -  \langle \Psi(t) \vert \hat{X} \vert \Psi(t) \rangle^2 \right ].
\label{eq:QFI}
\end{equation}
  We will be focusing on the LTA of the QFI, so to simplify the notation we will use a bar to represent the LTA, $\overline{\mathcal{O}} =\lim_{T\to \infty}\frac{1}{T} \int_0^T \mathcal{O}(t) dt$, where $\mathcal{O}(t)$ is some function of time $t$.  Expanding $\hat{X}$ in Eq.\ \eqref{eq:QFI} in its eigenbasis, $\sum_i x_i \vert x_i \rangle \langle x_i \vert$, the LTA of the QFI can be written as

\begin{equation}
\overline{F_Q}  = 4 \left [\sum_i x_i^2 \overline{P(x_i)} - \sum_{i,j} x_i x_j \overline{P(x_i)P(x_j)} \right ]
\label{eq:QFI2}
\end{equation}
where $P(x_i,t) = \vert \langle x_i \vert \Psi (t)\rangle \vert^2$ is the probability of finding the system in eigenstate $\vert x_i \rangle$ at time $t$.  

Our first approximation is to assume that there are no correlations between different points of the probability distribution at long times.  This means that the LTA of the second term in Eq.\ \eqref{eq:QFI2} can be factored, 

\begin{equation}
\overline{P(x_i)P(x_j)} \approx \overline{P(x_i)}\,\,\overline{P(x_j)}.
\label{eq:factor}
\end{equation}
Although this approximation fails when $x_i = x_j$, its effect on Eq.\ \eqref{eq:QFI2} is mitigated when the dynamics explores many eigenstates of $\hat{X}$. In that case, the sum is dominated by terms with $x_i \neq x_j$, since the number of these terms grows quadratically with the number of accessible states, while the number of diagonal terms $x_i = x_j$ only grows linearly. Therefore, the error introduced by neglecting the correlations becomes negligible when the number of accessible states is large (see Appendix \ref{app:error}) leading to the approximated QFI

\begin{equation}
\overline{F_Q}  \approx 4 \left [\sum_i x_i^2 \overline{P(x_i)} - \left (\sum_{i} x_i \overline{P(x_i)}\right )^2 \right ].
\label{eq:QFI3}
\end{equation}

\subsection{Mean-field and Semiclassical Approximations}

Furthermore, we assume that the quantum many-body system being analyzed has a classical limit in which there are conjugate variables akin to position and momentum.  This means that the dynamics can be described by the motion of a particle in an effective potential.  In integrable models where energy is conserved, the classical equations of motion can be combined into one

\begin{equation}
\left ( \dot{x} \right )^2 + V_\mathrm{eff} \left (x \right ) = 0
\label{eq:eom}
\end{equation}
where $x$ is the position variable and $V_\mathrm{eff}(x)$ is the effective potential energy.  The potential energy can also be written in terms of its roots

\begin{equation}
V_\mathrm{eff}(x) \propto \prod_{j=1}^{N_R} \left (x - r_j \right ),  
\label{eq:roots}
\end{equation}
so any change in the dynamics can be ascribed to the change in the structure or the number of roots.  

The quantum and classical behavior is linked with the semiclassical approximation that $\overline{P(x_i)}$ can be written in terms of a probability density function  which is inversely proportional to the speed of the classical particle 

\begin{equation}
\mathcal{P}(x)  =\frac{\mathcal{N}}{\vert \dot{x} \vert}
\label{eq:prob1}
\end{equation}
where $\mathcal{N}$ is a normalization constant.  The connection between the discrete eigenvalue $x_i$ and the continuous variable $x$ is given by $x = \left(x_i - \Delta_\mathrm{ave} \right )/\Delta_\mathcal{D}$ where $\Delta_\mathcal{D} = \left ( x_\mathcal{D} - x_1\right )/2$ is the half-span and $\Delta_\mathrm{\mathrm{ave}} = \left (x_\mathcal{D} + x_1\right )/2$ is the average of the minimum and maximum of the eigenvalues of $\hat{X}$, respectively, with $\mathcal{D}$ being the dimension of the Hilbert space. Assuming the eigenvalues are in ascending order, this transformation ensures that as the index $i$ increases from 1 to 
$\mathcal{D}$, the variable $x$ spans the interval $[-1,1]$. While $x$ is technically not continuous due to the finite spacing $\Delta x = x_{i+1} - x_i$, we assume $\frac{\Delta x}{\Delta_\mathcal{D}} \ll 1$, so the continuum approximation is valid.  The continuous probability density function approximates the LTA of the quantum distribution with the relation $\overline{P(x_i)} \approx \mathcal{P}(x_i) \frac{\Delta_x}{\Delta_\mathcal{D}}$. 
After combining Eqns.\ \eqref{eq:eom}, \eqref{eq:roots} and Eq.\ \eqref{eq:prob1} the probability density funciton is

\begin{equation}
\mathcal{P}(x) = \frac{\mathcal{N}}{\sqrt{\left [x(0)-x\right ]\prod_{j=2}^{N_R} \left (x - r_j \right )}}.
\label{eq:SCA}
\end{equation}
where $x(0)$ is the initial position and we have adopted the convention that $r_1 = x(0) > r_2>r_3...$.  Without loss of generality, here, and throughout the rest of the paper we take the initial momentum to be zero, $p(0) = 0$.  With the semiclassical approximation in Eq.\ \eqref{eq:SCA} the LTA of the QFI becomes

\begin{equation}
\overline{F_Q}  \approx 4 \underbrace{\Delta_\mathcal{D}^2\left [  \int_{r_i}^{x(0)} dx x^2 \mathcal{P}(x) - \left ( \int_{r_i}^{x(0)} dx x \mathcal{P}(x) \right )^2 \right ] }_{\equiv \sigma_X^2}
\label{eq:QFI4}
\end{equation}
where the bounds of integration $x(0)$ and $r_i$ are two of the roots of $V_\mathrm{eff}(x)$ that represent the classical turning points.

\section{Characterizing DQPTs with the QFI}

 In the following we use Eq.\ \eqref{eq:QFI4} to derive general expressions of the variance $\sigma_X^2$ (i.e., $\overline{F_Q}/4$) for two common types of DQPTs.
\\ \\
Type A: The potential has $N_R$ roots, however, only one pair of roots is needed to describe the dynamics for a given set of parameter values.  At a specific set of parameter values, the relevant pair abruptly changes.  This situation describes a particle trapped in a single well, but a transition occurs when the well suddenly changes shape.   The probability density function for this case is

\begin{equation}
\mathcal{P}_\mathrm{A}(x) = \frac{1}{\pi\sqrt{\left [x(0)-x\right ] \left [x - r \right ]}}
\label{eq:LTA2}
\end{equation}
where $\mathcal{N} = 1/\pi$ and the subscript of $r$ has been dropped with the understanding that $x(0) \geq x > r$.  The variance is then

\begin{equation}
\sigma_{X,\mathrm{A}}^2 = \frac{\Delta_\mathcal{D}^2}{8} \left [x(0) -r  \right ]^2.
\label{eq:sigmac}
\end{equation}
and the DQPT is marked by a sudden change in $r$.\\ \\
Type B: The potential has the form of a symmetric double-well.  This means there are four roots with $r_1 = -r_4= x(0)$ and $r_2 = - r_3 = r$.  At a specific energy the system transitions from a trapped phase where the dynamics is confined to one well to an untrapped phase where both wells are explored.  The transition occurs suddenly when the energy of the particle has enough to overcome the central barrier.  The probability density function for this case is

\begin{equation}
\mathcal{P}_\mathrm{B}(x) =\frac{\mathcal{N}}{\sqrt{\left [x(0)^2-x^2\right ] \left [x^2 -  r^2 \right ]}}
\label{eq:LTA3}
\end{equation}
where the normalization constant has been kept general because it depends on what phase the system is in.  The resulting variances for the trapped and untrapped phases are 

\begin{equation}
\sigma_{X,\mathrm{B1}}^2 =\left [  \frac{\Delta_\mathcal{D}^2x(0)^2}{K\left ( m^{-1} \right )^2}\right ]\left [  E\left ( m^{-1}  \right ) K\left ( m^{-1}  \right ) -\left ( \frac{\pi}{2}\right )^2 \right ]
\label{eq:sigmatr}
\end{equation}
and

\begin{equation}
\sigma_{X,\mathrm{B2}}^2 =\Delta_\mathcal{D}^2 x(0)^2 \left (\frac{1-m}{m}\right ) \left [ \frac{E \left ( \frac{m}{m-1}\right )}{K \left ( \frac{m}{m-1}\right )}-1\right ],
\label{eq:sigmautr}
\end{equation}
respectively.  The functions $K(y)$ and $E(y)$ are the complete elliptic integrals of the first and second kind, respectively, and $m = \frac{x(0)^2}{x(0)^2-  r^2}$.  In the trapped phase, $r$ is real, so $m>1$ while in the untrapped phase $r$ is imaginary, so $m<1$ marking $m_c=1$ as the critical point in the DQPT. 



%

%

Equations \eqref{eq:sigmac}, \eqref{eq:sigmatr} and \eqref{eq:sigmautr} are the main results of the paper.  Examining the latter two, we see that the quantity $\left [\sigma_X/(x(0) \Delta_\mathcal{D}) \right ]^2$ is universal in the sense that any DQPT from a trapped to an untrapped phase involving a symmetric double-well potential will follow these expressions, once the appropriate identification of the parameter $m$ is made.  This implies that $\overline{F_Q}/(x(0)\Delta_\mathcal{D})^2$ is also universal and in various limits of $m$, simplified general expressions can be derived.  For example, deep in the untrapped phase ($m\ll1$), the leading order behavior is $\overline{F_Q} \approx 2(1-m/8) $ while deep in the trapped phase ($m\gg1$),  $\overline{F_Q} \approx m^{-2}/8 $.  Near the critical point, as $m\to1^-$, the behavior is approximately   $\overline{F_Q} \approx 2\left [m-1+\frac{m+3}{\ln\left(\frac{16}{1-m}\right)} \right ]$, although the corresponding expression for $m\to1^+$ is less compact.  These expressions are system-independent and rely only on the DQPT being of Type B.


In general, the Heisenberg limit marks the largest possible value of the QFI, $F_Q^\mathrm{H} = 4\Delta_\mathcal{D}^2$, and comes from states with the largest variance of $\hat{X}$ which are cat states in the $\hat{X}$ eigenbasis, $\vert \mathrm{cat}\rangle = \left (\vert x_1\rangle+\vert x_\mathcal{D}\rangle \right )/\sqrt{2}$.  
We can then ask: what is the maximum LTA of the QFI for the conditions we consider?  Looking at Eqns.\ \eqref{eq:sigmac}, \eqref{eq:sigmatr} and \eqref{eq:sigmautr}, the largest variance is $\Delta_\mathcal{D}^2/2$ when $x(0) = 1$, leading to $\left (\overline{F_Q}\right )_\mathrm{max} = F_Q^\mathrm{H}/2 = 2\Delta_\mathcal{D}^2$.  The reason why the LTA of the QFI must be less than the Heisenberg limit is simply because the LTA of the probability distribution is nonzero between the edges of the Hilbert space, so the variance will always be less than that of the cat state.

\section{Descriptions of the Spinor-BEC and the LMG Model}

Here, we introduce two systems with which we will investigate the two types of DQPTs.  The first are the static \cite{zhang05,black07,liu09,kawaguchi12} and driven \cite{austin24,guan25} versions of the spinor-BEC and the second is the LMG model.  For each system, from the quantum many-body Hamiltonian we derive the mean-field effective potential and identify its roots as well as the critical point of the DQPT.

\subsection{Static and Driven Spinor-BEC}

 We begin with the static spinor-BEC which is a dilute gas of $N$ spin-1 bosons confined to an optical dipole trap \cite{kawaguchi12,kurn13}. We assume that the single-mode approximation can be applied, which is valid when the energy scale associated with spatial degrees of freedom is much larger than that of spin. Under this assumption, the spatial and spin degrees of freedom effectively decouple, and all spin components occupy the same spatial mode, $\phi(\bf{r})$.  This greatly simplifies the model, allowing one to write the Hamiltonian entirely in terms of the spin modes.  The Hamiltonian itself has three contributions, $\hat{H}_\mathrm{S} = \hat{H}_\mathrm{inel}+\hat{H}_\mathrm{el}+\hat{H}_\mathrm{Z}$ where

\begin{eqnarray}
\hat{H}_\mathrm{inel} &=& \frac{c}{N} \left (\hat{a}_0^\dagger \hat{a}_0^\dagger \hat{a}_{1} \hat{a}_{-1} + h.c. \right ), \nonumber \\
\hat{H}_\mathrm{el} &=& \frac{c}{N} \hat{a}_0^\dagger \hat{a}_0\left (\hat{a}_{1}^\dagger \hat{a}_1 + \hat{a}^\dagger_{-1} \hat{a}_{-1} \right ), \nonumber \\
\hat{H}_\mathrm{Z} &=& q\left (\hat{a}_{1}^\dagger \hat{a}_1 + \hat{a}^\dagger_{-1} \hat{a}_{-1} \right ).
\label{eq:ham}
\end{eqnarray}
The first and second terms correspond to inelastic spin-changing collisions and elastic spin-conserving collisions, respectively. The final term accounts for the quadratic Zeeman effect, which induces an energy shift in the $m_s = \pm 1$ spin components. In general, the Hamiltonian also includes terms that are linear and quadratic in the magnetization $\hat{M} = \hat{a}_1^\dagger \hat{a}_1 - \hat{a}_{-1}^\dagger \hat{a}_{-1}$. However, since the magnetization is conserved, $[\hat{M}, \hat{H}  ] = 0$, these terms only contribute a global phase to the dynamics of any state, which can be absorbed into a rotating frame.  The parameter $q$ represents the strength of the quadratic Zeeman shift, arising from an external magnetic field. The interaction strength $c$ characterizes the effective two-body interactions within the system. Assuming the gas is at a sufficiently low temperature, the interactions are dominated by $s$-wave scattering. Under the single-mode approximation, the parameter $c$ scales with the density squared of the gas.

To apply the analysis outlined in the previous section we follow Ref.\ \cite{zhang05} and assume that $N \gg 1$, so the mean-field approximation can be applied.  This becomes exact in the thermodynamic limit, $N \to \infty$.  The mean-field approximation amounts to replacing the mode operators with their coherent state expectation values, $\hat{a}_i \to \sqrt{N} \alpha_i$, which can be parameterized by a phase and an amplitude, $\alpha_i = \sqrt{\rho_i} e^{i \theta_i}$ where $i=-1,0,+1$.  Putting this substitution into $\hat{H}/N$ gives the mean-field Hamiltonian for the static spinor-BEC

%
\[
\begin{aligned}
H_\mathrm{mf}^\mathrm{S} &= c \rho_0 (1 - \rho_0) + c \rho_0 (1 - \rho_0) \cos(2\theta) \\
&\quad + q (1 - \rho_0).
\end{aligned}
\]
Here, $ \rho_i = n_i / N $ denotes the fractional population in spin state $i$ and $ \theta = \theta_0 - (\theta_1 + \theta_{-1})/2 $ is a relative phase variable. The mean-field Hamiltonian is expressed solely in terms of the zero mode due to the constraint of particle number conservation, $\sum_i \rho_i = 1$. Furthermore, without loss of generality, we have also set the mean-field magnetization, $M = \rho_1 - \rho_{-1}$, to zero which is conserved.  The variables $\rho_0$ and $\theta$ form a conjugate pair acting as position and momentum, respectively. Because we are assuming that the mean-field initial state is that of a displaced particle with zero momentum, the initial position is $\rho_0(0)$ while the initial momentum is $\theta(0) = 0$.  Applying Hamilton’s equations of motion, $\dot{\rho}_0 = -\partial H_\mathrm{mf}^\mathrm{S} / \partial \theta$, and using energy conservation, one obtains an equation of motion of the same form as Eq.\ \eqref{eq:eom} with the substitution $x \to \rho_0$.  The resulting effective potential of the static spinor-BEC is

\begin{eqnarray}
V_\mathrm{eff}^{\mathrm{S}}(\rho_0) &=& 4 c^2 \rho_0^2 (1-\rho_0 )^2- 4 \left \{ c\left [1-\rho_0(0) \right ] \rho_0(0) \right. \nonumber \\
&& \left.+\left [1-\rho_0(0) \right ]\left[q+c \rho_0(0) \right ] - (1-\rho_0)(q+c \rho_0 )\right \}^2 \nonumber \\
\label{eq:Veff1}
\end{eqnarray}
which has the following three roots:

\begin{eqnarray}
r_1 &=& \rho_0(0) \\
r_2 &=& 1-\rho_0(0) - \frac{q}{2c} \\
r_3 &=& \rho_0(0) - \frac{2c}{q} \rho_0(0) \left [1-\rho_0(0) \right ].
\end{eqnarray}
However, for any given value of $q$, only two of the roots lie within the physical range $0\leq \rho_0 \leq 1$ which means Eq.\ \eqref{eq:Veff1} is a potential for a Type A DQPT. When $q<q_c$, the relevant second root is $r=r_2$, while for $q>q_c$, it becomes $r=r_3$, where the critical value is given by $q_c=2c\left [1-\rho_0(0) \right ]$. This critical point marks a transition between two distinct dynamical phases: one where $\theta(t)$ evolves continuously through all values, and another where $\theta(t)$ remains confined to a finite interval. In both phases, $\rho_0(t)$ is confined to a finite interval, but the interval abruptly changes at the critical point.

Recently it was shown in an experiment that by periodically driving the width of the trapping potential of the spinor-BEC the spin interaction energy is also periodically modulated \cite{austin24,guan25}. In terms of the Fourier series, the interaction energy is $c(t) = \mathcal{G}_0 + \sum_k \mathcal{G}_k \cos(k\omega t)$ where $j\omega$ is the drive frequency and $\mathcal{G}_j$ is the amplitude of the $j^\mathrm{th}$ term in the series.  Under the near resonance condition $q \approx j\omega/2$, the periodic driving leads to an effective Hamiltonian  where the two spin interaction terms can be controlled independently: the inelastic term becomes $\hat{H}_\mathrm{inel} \to \frac{\mathcal{G}_j}{2N} \left (\hat{a}_0^\dagger \hat{a}_0^\dagger \hat{a}_1\hat{a}_{-1}+h.c. \right )$, while the elastic term becomes $\hat{H}_\mathrm{el} \to \frac{\mathcal{G}_0}{N} \hat{a}^\dagger_0\hat{a}_0 \left (\hat{a}_1^\dagger \hat{a}_1 + \hat{a}^\dagger_{-1} \hat{a}_{-1} \right )$.  There is also a new Zeeman shift, $\tilde{q} = q-j\omega/2$, however, since we are considering a symmetric effective potential we set $\tilde{q} = 0$.

Applying the mean-field treatment to the effective Hamiltonian gives the driven effective potential

\begin{eqnarray}
V_\mathrm{eff}^\mathrm{D}(\rho_0)/\mathcal{G}_0 &=&  -\eta^2 \rho_0^2 (1-\rho_0 )^2+ \left [ 2 \rho_0 \left (1-\rho_0 \right)\right. \nonumber \\
&&\left. - \left (2+\eta \right ) \rho_0(0) \left (1-\rho_0(0) \right ) \right ]^2
\label{eq:Veff2}
\end{eqnarray}
where $\eta = \mathcal{G}_j/\mathcal{G}_0$.  This potential is a symmetric double-well which means it describes a Type B DQPT with roots

\begin{eqnarray}
r_1 &=& \rho_0(0) \\
r_2 &=& \frac{1}{2} \left[ 1+ \sqrt{1-4\left (\frac{2+\eta}{2-\eta}\right )\rho_0(0)\left (1-\rho_0(0) \right )}\right ] \\
r_3 &=& \frac{1}{2} \left[ 1- \sqrt{1-4\left (\frac{2+\eta}{2-\eta}\right )\rho_0(0)\left (1-\rho_0(0) \right )}\right ] \\
r_4 &=& 1-\rho_0(0)
\end{eqnarray}
and the critical point of the DQPT can be found by calculating when $r_2$ and $r_3$ become complex giving $\eta_c = \frac{2\left [1-2\rho_0(0) \right ]^2}{1+4\rho_0(0) - 4 \rho_0(0)^2}$.  It is important to note that the roots are not in the symmetric form used in Eq.\ \eqref{eq:LTA3} because the double-well potential is shifted away from the origin, however, the transformation $r_i \to 2(r_i - 1/2)$  shifts the center to the origin. After this change of variables, the roots take the desired symmetric form: $r_1 = -r_4 = 2\rho_0(0) - 1$ and $r_2 = -r_3 =\sqrt{1-4\left (\frac{2+\eta}{2-\eta}\right )\rho_0(0)\left (1-\rho_0(0) \right )}$.

\subsection{LMG Model}

The second system we investigate is the Lipkin-Meshkov-Glick (LMG) model, which describes $N$ two-state bosons interacting via all-to-all couplings and has been studied in many systems including trapped ions \cite{jurcevic17,zhang17}, cavity-QED \cite{muniz20} and cold atoms \cite{albiez05,zibold10,muessel15}.  It is described by the Hamiltonian

\begin{equation}
\hat{H}_\mathrm{LMG} = \frac{2 \chi}{N}\hat{J}_z^2-2\Omega \hat{J}_x,
\end{equation}
where $\hat{J}_\alpha = \sum_i \hat{\sigma}_i^\alpha/2$ are collective spin operators and $\hat{\sigma}_i^\alpha$ are Pauli matrices of the $i^\mathrm{th}$ boson representing the two states and $\alpha = x,y,z$.  In terms of the bosonic modes, the collective spin operators are $\hat{J}_z = \left(\hat{a}_R^\dagger \hat{a}_R - \hat{a}_L^\dagger \hat{a}_L\right)/2$ and $\hat{J}_x = \left (\hat{a}_R^\dagger \hat{a}_L + \hat{a}_L^\dagger \hat{a}_R \right )/2$ where $L/R$ label the two modes.  The parameters $\chi$ and $\Omega$ control the strength of the all-to-all interactions and tunneling between the two states, respectively.  

After applying a mean-field approximation, the effective potential can be expressed as \cite{raghavan99}, 

\begin{equation}
V_\mathrm{eff}^\mathrm{LMG}(z) = z^2 - 1 + \left [\frac{\chi}{2} z^2  -E_0\right ]^2
\label{eq:VeffLMG}
\end{equation}
where $E_0 = \chi z(0)^2/2 - \Omega \sqrt{1-z(0)^2}\cos\phi(0)$, $z = 2n/N$ and $n$ is an eigenvalue of $\hat{J}_z$.  The phase angle $\phi$ is the conjugate variable to $z$, and since $z$ plays the role of the position coordinate, $\phi$ can be interpreted as the corresponding momentum. Following the previous choice of initial conditions with zero initial momentum, we set $\phi(0) = 0$. Under this condition, the classical turning points, or roots, of the effective potential are given by $r_1 = -r_4 = z(0)$ and $r_2 = -r_3 = r$, where 

\begin{equation}
r = \tfrac{\sqrt{2}}{\chi} \sqrt{\chi E_0-1-\sqrt{1-2\chi E_0+\chi^2}},
\end{equation}
so Eq.\ \eqref{eq:VeffLMG} is also a potential for a Type B DQPT. The critical value of $\eta$ is found with the condition $r = 0$ which gives $\chi_c = 2\left (1+\sqrt{1-z(0)^2} \right )/z(0)^2$.  The eigenvalue $n$ takes integer values in the range $-N/2 \leq n \leq N/2$, so $\Delta_\mathcal{D} = N/2$, $\Delta_\mathrm{ave} = 0$ and $x(0) = z(0)$. 

\section{Numerical Results}

\subsection{Long-Time Averaged Behavior of the Spinor-BEC}

\begin{figure}[t]
\centering
\includegraphics[width=\columnwidth]{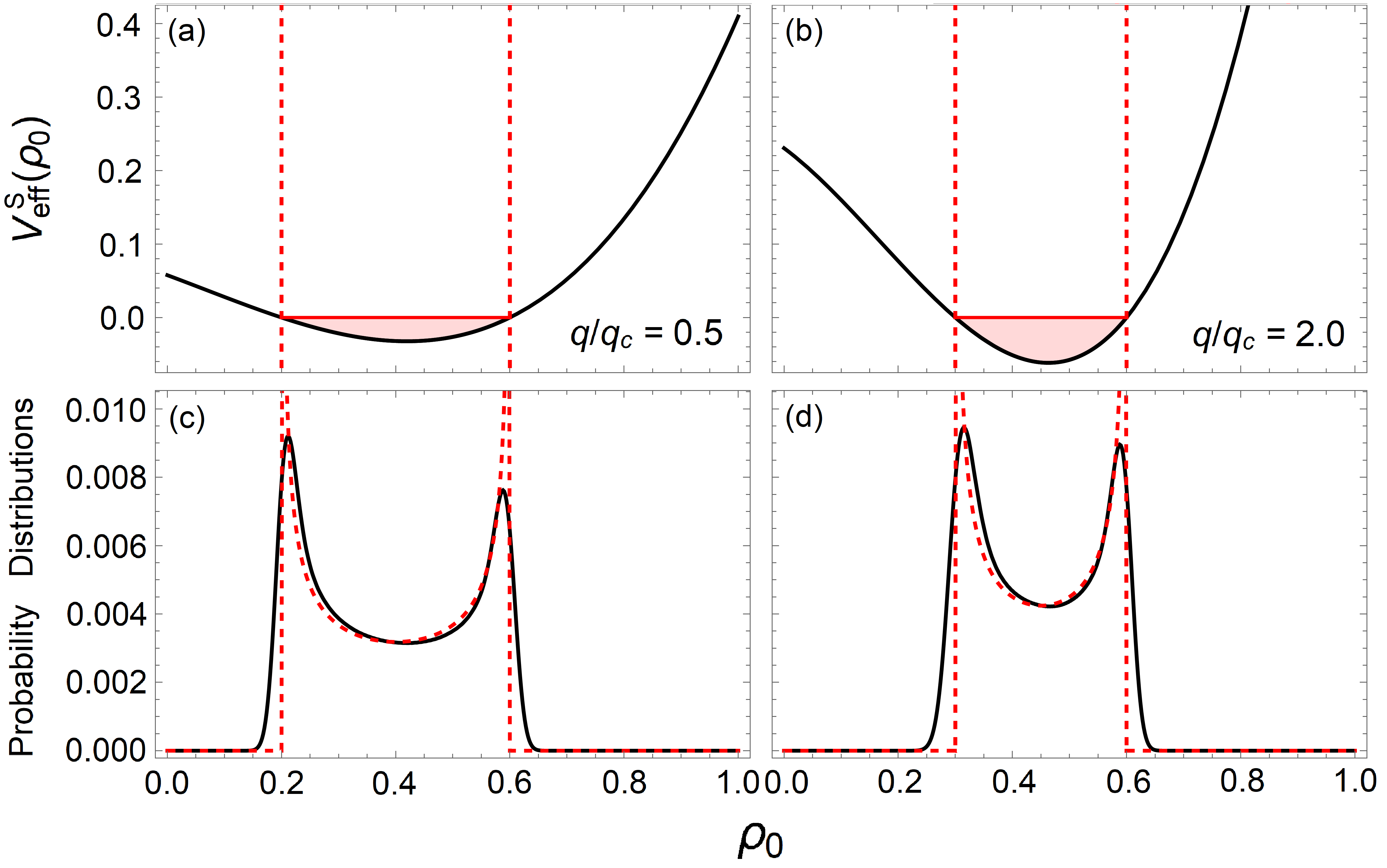}
\caption{Effective potential energy from Eq.\ \eqref{eq:Veff1} and LTA of the probability distribution as a function of $\rho_0$ for a Type A DQPT.  (a) and (b) show the effective potential for $q<q_c$ and $q>q_c$, respectively.  The vertical red dashed lines mark the locations of the turning points (roots) and the shaded parts highlight the classically accessible region of the potential well.  (c) and (d) show the quantum many-body (solid black)  and semiclassical (dashed red) results of the LTA of the probability distribution as a function of $\rho_0$.  The number of particles in the quantum simulation is $N = 1000$ with $c=1$ and the time average is taken from $2\times 10^4$ time steps of size $\Delta t = 1$.}
\label{fig:cubicpic}
\end{figure}

Here, we explore the validity of the semiclassical approximations in the spinor-BEC system. For the semiclassical analysis to hold, the quantum dynamics must remain largely confined between the classical turning points. As a result, we expect the agreement to deteriorate in regimes where quantum tunneling becomes significant. This requirement also places constraints on the choice of the initial quantum state, which must be sufficiently localized to resemble a classical particle. To satisfy this, we initialize the system in a coherent state  $\vert \rho_0, \theta \rangle = \vert \rho_0(0), 0\rangle$ (see Appendix \ref{sec:coh} for derivation), ensuring it closely matches a classical particle starting at position $\rho_0(0)$ with zero initial momentum. 

We compare $\mathcal{P}_\mathrm{A}(x)\frac{\Delta_x}{\Delta_\mathcal{D}}$ with $\overline{P(x_i)}$ in Fig.\ \ref{fig:cubicpic} which displays $V_\mathrm{eff}^\mathrm{S}(\rho_0)$ for $q<q_c$ and $q>q_c$ in panels (a) and (b), respectively.  The vertical lines indicate the positions of the classical turning points (roots). Panels (c) and (d) present the LTA of the probability distributions obtained from quantum many-body dynamics (black) alongside the semiclassical prediction (dashed, red). The two show excellent agreement across most of the domain, with noticeable discrepancies only near the turning points, where the classical prediction diverges due to the vanishing velocity there.

\begin{figure}[t]
\centering
\includegraphics[width=\columnwidth]{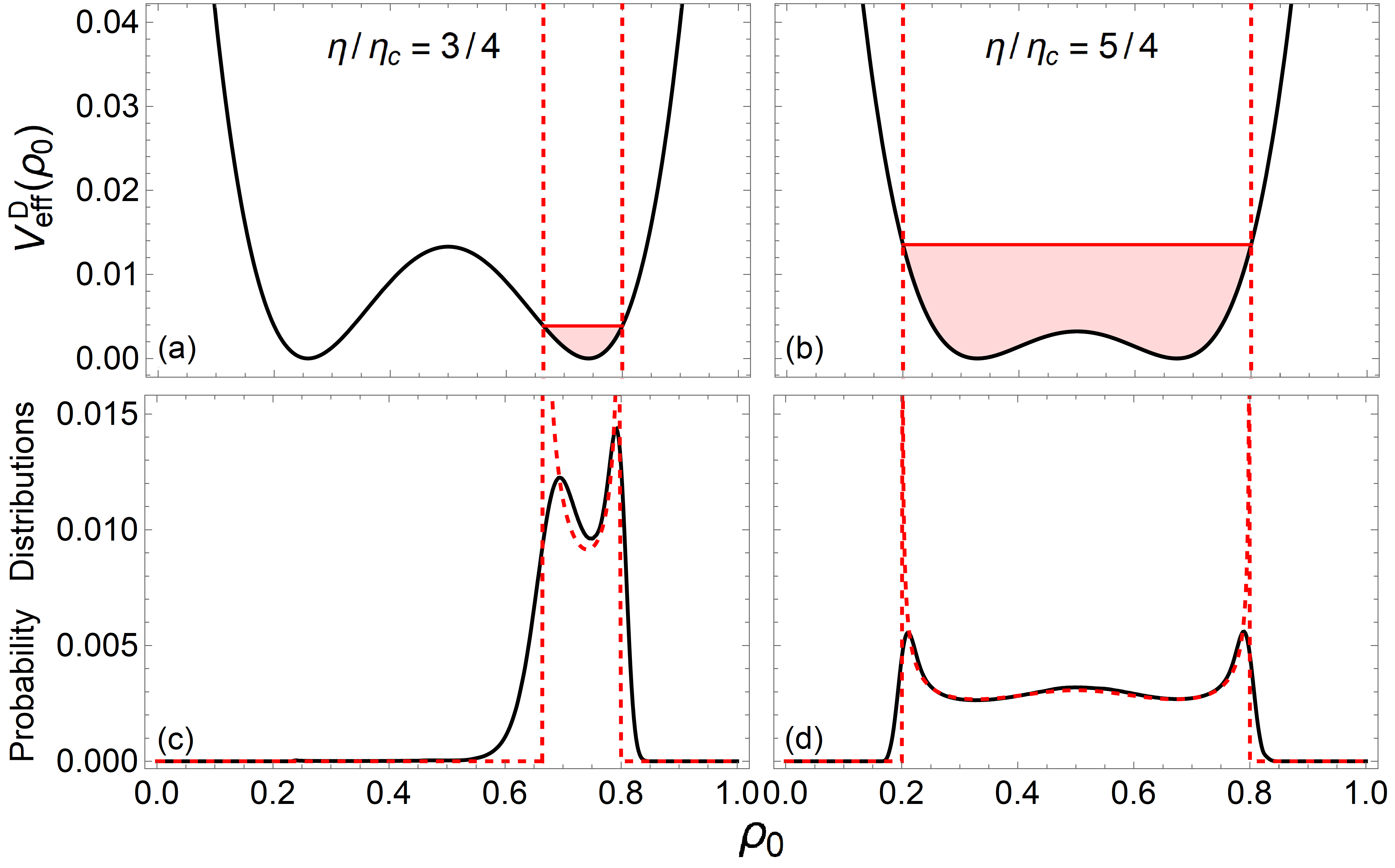}
\caption{Effective potential from Eq.\ \eqref{eq:Veff2} and LTA of the probability distribution as a function of $\rho_0$ for a Type B DQPT.  (a) and (b) show the effective potential of the trapped and untrapped phases, respectively.  (c) and (d) show the quantum many-body (solid black) and semiclassical (dashed red) results of the LTA of the probability distribution as a function of $\rho_0$.  The number of particles is $N = 1000$ and the time average is taken from $2\times 10^4$ time steps of size $\Delta t = 1$.}
\label{fig:quarticpic}
\end{figure}

Figure \ref{fig:quarticpic} presents the same set of plots as those shown in Fig.\ \ref{fig:cubicpic}, but for $\mathcal{P}_\mathrm{B}(x)\frac{\Delta_x}{\Delta_\mathcal{D}}$ and $V_\mathrm{eff}^\mathrm{D}(\rho_0)$. Panels (a) and (b) correspond to the trapped and untrapped phases, respectively, with the shaded region indicating the classically accessible part of the potential. Panels (c) and (d) compare the quantum many-body calculations (black) with the classical predictions (red dashed). As before, there is excellent overall agreement between the two calculations, however, the agreement is noticeably worse in the trapped case. This discrepancy arises because the initial coherent state has a finite width, and when the size of the trapped region becomes comparable to this width, quantum fluctuations become significant and the mean-field approximation begins to fail.  

\subsection{Quantum Fisher Information\label{sec:QFI}}

\begin{figure}[t]
\centering
\includegraphics[width=\columnwidth]{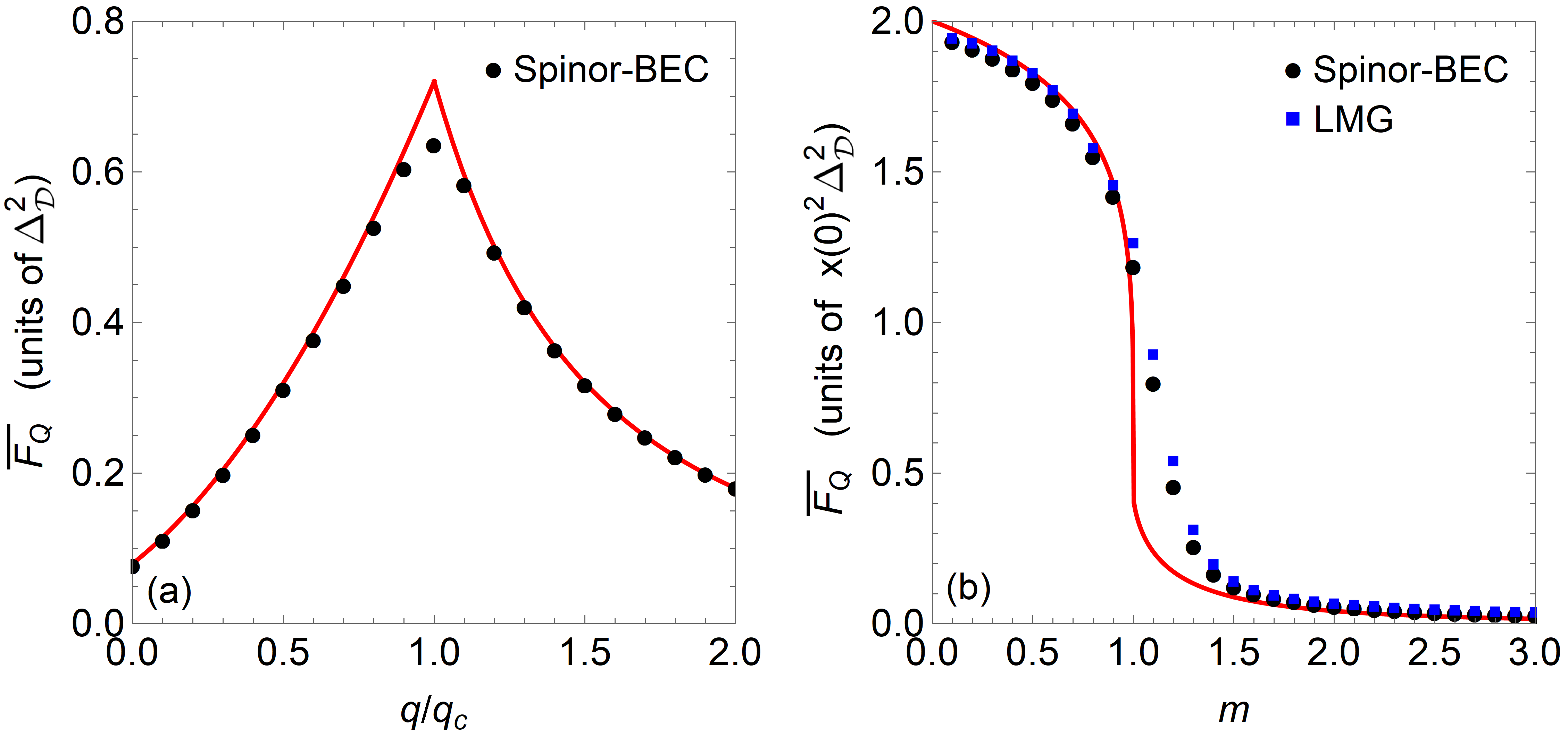}
\caption{LTA of the QFI for Type A and Type B DQPTs.  Data points are calculated from Eq.\ \eqref{eq:QFI} while red curves are calculated from Eq.\ \eqref{eq:QFI4}.  (a) Data for the static spinor-BEC as a function of the scaled quadratic Zeeman shift, $q/q_c$.  The initial state is a coherent state centered at $(\rho_0(0),\theta(0)) = (0.6,0)$.  (b) Data for the driven spinor-BEC (black circles) and the LMG model (blue squares) as a function of $m$.  The initial states are coherent states centered at $(\rho_0(0),\theta(0)) = (0.8,0)$ and $(z(0),\phi(0)) = (0.6,0)$ for the spinor-BEC and LMG model, respectively. The number of bosons for the spinor-BEC is $N = 1000$ and for the LMG model is $N = 500$.  The LTA is taken over $10^4$ time steps of size $\Delta t = 10$.}
\label{fig:CQQFI}
\end{figure}

In the spinor-BEC the position variable of the effective potential is $\rho_0$, so we choose $\hat{X} = \hat{n}_0=\hat{a}_0^\dagger\hat{a}_0$.  The analytic expressions for the LTA of the QFI for the Type A DQPT are calculated from Eq.\ \eqref{eq:sigmac} using the roots of $V_\mathrm{eff}^\mathrm{S}(\rho_0)$ and are $\overline{F_Q^{\leq}}\approx2 \Delta_\mathcal{D}^2\left [\rho_0(0) + (q-q_c)/2 \right ]^2$ for $q\leq q_c$ and $\overline{F_Q^{\geq}}\approx2\Delta_\mathcal{D}^2 \left [ \frac{q_c \rho_0(0)}{q}\right]^2$ for $q \geq q_c$. 
 Figure \ref{fig:CQQFI}(a) compares these results (red curves) with the quantum many-body calculations from Eq.\ \eqref{eq:QFI} (black circles). Both the QFI and the analytic approximation are normalized by $\Delta_\mathcal{D}^2 = (N/2)^2$ and plotted as a function of the scaled Zeeman shift $q/q_c$.  Despite the various approximations used in the analytic derivation, the agreement with the numerical results is very good, except near the critical point, where quantum fluctuations smooth out the sharp kink predicted by the analytic expressions.  

 In Figure \ref{fig:CQQFI}(b) we make the same comparison for the driven case where the system undergoes a Type B DQPT from untrapped ($m<1$) to trapped ($m>1$).  The analytic results (red curves) are the LTA of the QFI using Eqns.\ \eqref{eq:sigmatr} and \eqref{eq:sigmautr} while the quantum many-body results are  again computed from Eq.\ \eqref{eq:QFI} (black circles). The QFI is scaled by $\Delta_\mathcal{D}^2x(0)^2$, and we find very good agreement across the full range of $m$, with the exception of the critical point at $m_c=1$, where deviations arise due to finite-size effects.  We also plot the QFI of the LMG model in Fig.\ \ref{fig:CQQFI}(b) (blue squares) using the roots from $V_\mathrm{eff}^{\mathrm{LMG}} (z)$ and with $\hat{X} = \hat{J}_z$ since $z$ serves as the position variable. We find almost exact agreement with the spinor-BEC data showing the universality of the scaled QFI.  Notably, the number of bosons in the spinor-BEC is double that of the LMG model, so that each system has the same Hilbert space dimension since $\mathcal{D}_\mathrm{S-BEC} = N/2+1$ and $\mathcal{D}_\mathrm{LMG} = N+1$.  Therefore, both systems suffer from the same finite size effects, so the slight difference in their data can be attributed to the different sizes of initial coherent states with the spinor-BEC initial state being slightly narrower.  It is expected that in the thermodynamic limit, $N\to\infty$, the data of both systems will collapse to the semiclassical result. This  can be partially understood by considering the properties of the initial coherent states. In each system, their widths  scale as $N^{-1/2}$ relative to the size of the Hilbert space. Therefore, as $N$ increases, the coherent states become increasingly localized, effectively approaching that of a classical point particle.


\section{Conclusions}

We have investigated DQPTs in systems with well-defined classical limits, focusing on the spinor-BEC and the LMG model. The LTA of the QFI was able to diagnose the DQPTs showing abrupt changes at the critical point. Two distinct types of DQPTs were analyzed: one in which the shape of a single potential well changes discontinuously, and another in which the system transitions between trapped and untrapped phases in a symmetric double-well potential. In the latter case, the semiclassical framework allowed us to show that the QFI exhibits universal features that apply across different physical models.

Although we focused on a single-well and a symmetric double-well potential, our methods can be applied more generally to potentials with any number of wells or asymmetries.  However, the analytic expressions quickly become cumbersome, see page 120 in Ref.\ \cite{byrd71} for the case of four \textit{distinct} roots which is applicable in the case of an asymmetric double-well potential.  In the driven spinor-BEC the asymmetry can come from a nonzero quadratic Zeeman shift whereas in the LMG model it can come from an energy imbalance between the two modes brought on by a term proportional to $\hat{J}_z$ in the Hamiltonian. 

\appendix

\section{Coherent states for the spinor-BEC and LMG models\label{sec:coh}}

Here we outline our calculations of the coherent states in both the spinor-BEC and the LMG model.  
\\ \\
For the spinor-BEC the coherent state is \cite{jie19}

\begin{equation}
    \vert \alpha_1,\alpha_0,\alpha_{-1} \rangle = \frac{1}{\sqrt{N!}} \left(\alpha_1 \hat{a}_1^\dagger + \alpha_0 \hat{a}_0^\dagger + \alpha_{-1} \hat{a}_{-1}^\dagger\right )^N \vert\mathrm{vac}\rangle
\end{equation}
where $\vert \mathrm{vac} \rangle$ denotes the vacuum state.  Applying the creation operators to the vacuum results in the expression

\begin{eqnarray}
    \vert \alpha_1,\alpha_0,\alpha_{-1} \rangle &=& \sum_{n_1=0}^N\sum_{n_0=0}^N\sum_{n_{-1}=0}^N\sqrt{\frac{N!}{n_1!n_0!n_{-1}!}} \nonumber \\ && \times\alpha_1^{n_1}\alpha_0^{n_0}\alpha_{-1}^{n_{-1}} \vert n_1,n_0,n_{-1}\rangle \nonumber \\
    &=& \sum_{n_0 = 0}^N \sqrt{\frac{N!}{n_0! [(N/2 - n_0/2)!]^2}} \nonumber \\
    &&\times \alpha_1^{N/2-n_0/2} \alpha_0^{n_0} \alpha_{-1}^{N/2-n_0/2} \vert n_0 \rangle
\end{eqnarray}
where in the second equation the constraints of particle number conservation $n_1+n_0 + n_{-1} = N$ and magnetization conservation $M=n_1 - n_{-1} $ have been applied.  In the paper we have $M = 0$, so $n_1 = n_{-1}$.  The coherent state coefficients 

\begin{equation}
    c_{n_0} = \sqrt{\frac{N!}{n_0! [(N/2 - n_0/2)!]^2}} \alpha_1^{N/2-n_0/2} \alpha_0^{n_0} \alpha_{-1}^{N/2-n_0/2} 
\end{equation}
can have ratios of very large numbers which can lead to numerical precision issues in some computational software. For this reason, we choose to calculate them using a recursion relation which deals with ratios of smaller numbers.  We choose to start with the $n_0 = N$ coefficient, $c_N = \alpha_0^N$, and work our way down, so the recursion relation ends up being

\begin{equation}
    c_{n_0 - 2} =2 c_{n_0} \frac{\sqrt{n_0(n_0-1)}}{N-n_0+2} \frac{\alpha_1 \alpha_{-1}}{\alpha_0^2}.
\end{equation}
Using the relation $\alpha_i =\sqrt{\rho_i} e^{i\theta_i}$ and the conservation constraints, $\rho_1 + \rho_0 + \rho_{-1} = 1$ and $\rho_1 = \rho_{-1}$, the coefficients become

\begin{equation}
    c_{n_0 - 2}=2 c_{n_0} \frac{\sqrt{n_0(n_0-1)}}{N-n_0+2}  \left (\frac{1-\rho_0}{2\rho_0}\right ) e^{-2i\theta}
\end{equation}
where $\theta = \theta_0 - (\theta_1 + \theta_{-1})/2$.  We note that when $N$ is even, the index $n_0$ takes even values since the spin mixing term in the Hamiltonian either annihilates or creates bosons in pairs in the zeroth mode.  The coherent state is then

\begin{equation}
\vert \rho_0, \theta \rangle=\mathcal{N}\sum_{n_0=0}^{N} c_{n_0} (\rho_0,\theta) \vert n_0\rangle
\end{equation}
where we have made it explicit that the coefficients are functions of the mean-field variables $\rho_0$ and $\theta$ and $\mathcal{N} = \left (\sum_{n_0=0}^{N} \vert c_{n_0} (\rho_0,\theta) \vert^2 \right )^{-1/2}$ is the normalization constant.\\ \\
For the LMG model the coherent state is similar to the spinor-BEC, but with two modes 

\begin{equation}
    \vert \alpha_R,\alpha_L \rangle = \frac{1}{\sqrt{N!}} \left(\alpha_R \hat{a}_R^\dagger + \alpha_L \hat{a}_L^\dagger \right )^N \vert\mathrm{vac}\rangle
\end{equation}
which are labeled $R$ and $L$.  Once again, applying the creation operators to the vacuum gives

\begin{eqnarray}
    \vert \alpha_R,\alpha_L \rangle &=& \sum_{n_R=0}^N\sum_{n_L=0}^N\sqrt{\frac{N!}{n_R!n_L!}} \nonumber \\ && \times\alpha_R^{n_R}\alpha_L^{n_L} \vert n_R,n_L\rangle \nonumber \\
    &=& \sum_{n = -N/2}^{N/2} \sqrt{\frac{N!}{(N/2+n)! (N/2 - n)!}} \nonumber \\
    &&\times \alpha_R^{N/2+n}\alpha_L^{N/2-n} \vert n \rangle
\end{eqnarray}
where particle number conservation $n_R+n_L = N$ is used as well as the new index $n = (n_R - n_L)/2$.  Starting with the $n = N/2$ coefficient, $d_{N/2} = \alpha_R^N$, the recursion relation is

\begin{equation}
    d_{n-1} =d_n \sqrt{\frac{N/2+n}{N/2-n+1}} \frac{\alpha_L}{\alpha_R} .
\end{equation}
Using the relation $\alpha_{R/L} = \sqrt{1\pm z}\,e^{i \phi_{R/L}}$ the coefficients become

\begin{equation}
    d_{n-1} =d_n \sqrt{\frac{N/2+n}{N/2-n+1}} \sqrt{\left (\frac{1-z}{1+z}\right )}e^{-i\phi} .
\end{equation}
where $\phi = \phi_R - \phi_L$.  The coherent state is then

\begin{equation}
    \vert z,\phi \rangle = \mathcal{N} \sum_{n=-N/2}^{N/2} d_n(z,\phi) \vert n\rangle
\end{equation}
where $\mathcal{N} = \left (\sum_{n=-N/2}^{N/2} \vert d_{n} (z,\phi) \vert^2 \right )^{-1/2}$.

\section{Errors in Factoring the Long-time Average of the Probability Distributions\label{app:error}}

\begin{figure}[t]
\centering
\includegraphics[width=\columnwidth]{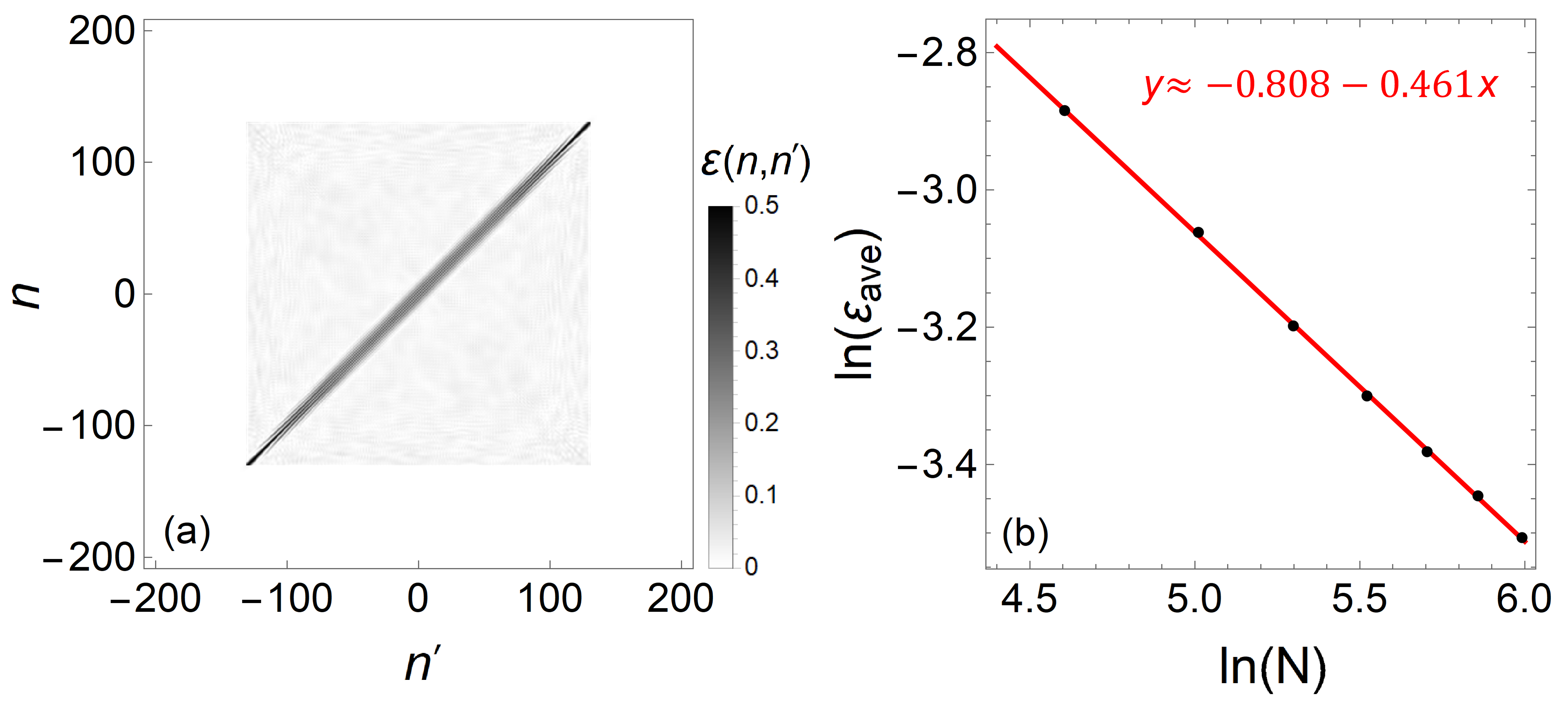}
\caption{Relative error and average relative error of the factorization approximation used in Eq.\ \eqref{eq:factor}.  (a) Density plot of $\varepsilon(n,n^\prime)$ for the LMG model for $N = 400$. (b) Log-log plot of $\varepsilon_\mathrm{ave}$ vs. $N$ (black circles) along with a linear fit (red line) where the range of $N$ values is $100\leq N \leq400$.  In both plots, $\chi = 5$, the initial state is a coherent state centered at $z(0) = 0.6$, $\phi(0)=0$ and the long-time average is taken over 20000 time steps of size $\Delta t = 1$.}
\label{fig:errorRow2}
\end{figure}

A key approximation underlying much of our analysis is the factorization of the LTA of the probability distributions, as introduced in Eq.\ \eqref{eq:factor}. To quantify the accuracy of this approximation we define the relative error as

\begin{equation}
\varepsilon(n,n^\prime) = \left \vert 1-\frac{\overline{P(n)}\,\overline{P(n^\prime)}}{\overline{P(n)P(n^\prime)}} \right \vert
\end{equation}
where $n$ and $n^\prime$ are eigenvalues of $\hat{J}_z$.  Figure \ref{fig:errorRow2}(a) displays a density plot of $\varepsilon(n,n^\prime)$ for the untrapped phase in the LMG model. As expected, the largest deviations arise from the diagonal elements ($n=n^\prime$), while the off-diagonal contributions are typically an order of magnitude smaller.

To evaluate how the approximation improves with system size, we compute the average relative error within the classically allowed region,

\begin{equation}
    \varepsilon_\mathrm{ave} = \frac{1}{A} \sum_{n,n^\prime=n_\mathrm{min}}^{n_\mathrm{max}} \varepsilon(n,n^\prime).
\end{equation}
For the initial position of $z(0)$, the classically allowed region is in the range $-z(0)\leq z \leq z(0)$, so  the maximum and minimum values of $n$ are approximately $n_\mathrm{max} \approx \lfloor{ N z(0)/2\rfloor}$ and $n_\mathrm{min} \approx -\lceil{ N z(0)/2\rceil}$ while $A = (n_\mathrm{max} - n_\mathrm{min})^2 \approx N^2 z(0)^2$.  Figure \ref{fig:errorRow2}(b) shows a plot of $\ln\varepsilon_\mathrm{ave}$ as a function of $\ln N$. The red line is a linear fit with a slope of approximately -0.461, indicating that $\varepsilon_\mathrm{ave} \propto N^{-0.461}$. This confirms that the factorization error decreases with increasing system size.  We find similar $N$ scaling of the relative error in the spinor-BEC system.

\bibliography{bib_MQC} \label{References}

\end{document}